\documentclass[pra,showpacs,floatfix,amsmath,amsfonts,twocolumn]{revtex4}
\usepackage{dcolumn}
\usepackage{bm}
\usepackage{epsfig}
\usepackage{color}
\usepackage{graphics}
\usepackage{amssymb}

\begin{document}

\title{Nonunitary Newtonian Gravity Makes Semiclassical Gravity Viable for All Practical Purposes }

\author{Sergio De Filippo}

\affiliation{Dipartimento di Fisica ``E.R. Caianiello'', and INFN
- Gruppo Collegato di Salerno, Universit\'a di Salerno, Via Giovanni
Paolo II, 84084 Fisciano (SA), Italy}
\date{\today}
\affiliation{{\color{blue}e-mail: sdefilippo@sa.infn.it}}
\begin{abstract}
Semiclassical Gravity replaces the energy momentum tensor $T_{\mu \nu}$ with its expectation value $\langle T_{\mu \nu}\rangle$ in Einstein equations, this making the Einstein tensor $G_{\mu \nu}$ a classical variable thus avoiding the quantization of the gravitational field. Jointly with the Copenhagen interpretation of QM, Semiclassical Gravity gives rise to cases of Superluminal Communication. It is shown in a specific instance that a model of Nonunitary Newtonian Gravity avoids Superluminal Communication by decohering linear superpositions of macroscopically distinct states, namely Schrodinger's cats.
\end{abstract}

\maketitle

\section{Introduction}
The two weak points of the von Neumann setting of QM, the macroscopic measurement apparatus and the vague and subjective notion of coarse graining entropy, may both be addressed and hopefully overcome by a nounitary modification of quantum dynamics \cite{mypapers1,entropy1,entropy2,entropy3}. The natural way to get a nonunitary evolution law is to add hidden degrees of freedom with a unitary evolution of the metasystem = physical + hidden, and get the nonunitary evolution of the physical system once the hidden degrees are traced out just like for open systems once the environment is traced out.
This in particular suggests the possibility of looking for the quantum foundations of thermodynamic entropy as entanglement entropy of physical degrees of freedom with the hidden ones \cite{mypapers1,entropy1,entropy2,entropy3}.
 One should at least require thermal equilibrium between the physical system and the hidden one:
\begin{equation}
\frac{1}{T}= \frac{d S}{d E_p}=\frac{d S}{d E_h}=\frac{d S}{d E_p}\left(\frac{d E_h}{d E_p}\right)^{-1}, \nonumber
\end{equation}
where $E_p$ and $E_h$ denote respectively the energy of the physical system and hidden one, and the entropy $S$ is one and the same for both systems for being the entanglement entropy of a bipartite system.

The above equation implies that $E_h$ and $E_p$ differ at most by a constant and, since this has to be so for any physical system, an almost unescapable conclusion is that every physical system must have an exact copy entangled with itself and with a symmetry constraint on the allowed states for the exchange of physical and hidden degrees of freedom.  The constraint plays a crucial role in order to eliminate the arbitrariness in considering some degrees of freedom as non observable, since in any constrained theory, like for instance Gubta-Bleuler's, the observable algebra is a subalgebra of the original dynamical algebra, as the original algebra does not have a faithful representation in the constrained state space. And, if one is trying to reconcile quantum physics with gravity, and in particular with black hole information puzzle, it is only natural to assume that the two systems get entangled through gravitational interaction, as it is the case for instance in \cite{mypapers1,mypapers2} where Nonunitary Newtonian Gravity (NUNG) is treated . And already in the sixties F. Karolihazy hinted that gravity could modify quantum dynamics inducing decohence for center of mass of macroscopic bodies, starting from the fuzziness of space time \cite{karolyhazy,frenkel}. Karolyhazy's idea was further developed within the realm of spontaneous collapse models (SCM) \cite{ penrose}.

In Semiclassical Gravity (SCG) the Einstein equations are written as
\begin{equation}
G_{\mu \nu} =8 \pi G \langle T_{\mu \nu} \rangle \nonumber
\end{equation}
where the energy momentum tensor $T_{\mu \nu}$ is replaced by its expectation value and
$$
G_{\mu \nu}=R_{\mu \nu} - \frac 12 g_{\mu \nu} R $$
where $R_{\mu \nu}, R$ respectively are the Ricci tensor and scalar, $G$ is the Newton constant and $g_{\mu \nu}$ the metric tensor. The replacement of the energy momentum tensor, that is of a quantum operator, with its expectation value allows to keep gravity classical without quantizing it \cite{SCGkibble,SCGMattingly,SCGKiefer,SCGHu}. This corresponds in the Newtonian limit to consider the expectation of the mass density as the source of gravitation.

In  section II a specific instance of Superluminal Communication (SLC) ensuing from SCG is described and in section III we see how NUNG avoids de facto this inconsinstency of SCG.

Of course Newtonian gravity as any instantaneous action at a distance interaction implies in principle SLC; the instance of SLC presented here \cite{pearleandsquires} - as other instances of SLC \cite{tilloyLDiosiPRD2016e2017} - obviously does not refer to this aspect of the Newtonian gravity but to the lack of entanglement due to SCG between the position of the (macroscopic) source of gravity and the particles acted upon by gravity. We furthermore do not commit to SCG  at fundamental level whence SN is got.

It should also be said that despite in the standard derivation of the newtonian limit of the SCG the Schrodinger Newton (SN) equation is obtained, the non-linear term will not be interesting for the following because its relevant effect would be to further slow the spreading of the two localized states already sufficiently slow in ordinary MQ for the center of mass of macroscopic bodies. It must also be said that the SN can be obtained as the mean field limit of a non-unitary theory such as the NUNG \cite{gr-qc}.

\section{Superluminal Communication from Semiclassical Gravity}
A macroscopic ball of radius ${\bold{r}}$ is in a state that is linear supersposition of two wawefunctions centered in points at macroscopic distance from each other along x axis, $|+dx\rangle$ and $|-dx\rangle$, and this state is entangled with a two state system $|+\rangle$, $|-\rangle$ at large distance in such a way that the entangled state is
\begin{equation}
|\psi> = |+dx\rangle |+\rangle + |-dx\rangle |-\rangle. \nonumber
\end{equation}
Along the z axis a particle ray is directed between the two positions of the ball, namely through x=0 and in absence of a measurement act on the far away two state system, the particles of the ray do not feel any force in $x$ direction from the ball as the mass distribution is symmetrical but, as soon as the two state system is forced to go in one of the two states by an act of measurement, the particle ray gets deviated in one of the two directions and a particle detector beyond the ball along the $z$ axis sees an interruption of detection and this amounts to an instantaneous communication between the site of the two state system and the particle detector \cite{pearleandsquires}. This would realize the longstanding Einstein's dream to prove that QM implies the possibility of SLC.

\section{Nonunitary Newtonian Gravity avoids Superluminal Communication in Semiclassical Gravity}

While up to now, in order to get SCG free of inconsistencies due to the possibility of SLC, several authors had recourse to spontaneous collapse models (SCM), here we refer to NUNG, that is a model of gravitational decoherence in terms of the density operators; at variance with SCM, it does not need phenomenological parameters or cutoff procedures and furthermore not only it has a natural special relativistic extension \cite{SDFandfilippomaimonephyslettb} but it was also derived as the Newtonian limit of a general relativistic model \cite{mypapers1}.

As described thoroughly in \cite{mypapers1,mypapers2}, within NUNG a macroscopic ball and its hidden copy form a bound state that at zero temperature is in the ground state with zero point energy and it is possible to have a center of (meta)-mass wawefunction that, though being fairly localized, has such a slow spreading to be considered stationary. Only the relative motion gives rise to relevant time dependence. In the following, like in \cite{mypapers1,mypapers2}, we call metastate the state of the composite metasystem physical + hidden balls, whereas the physical state is obtained by tracing out the hidden ball.
Furthermore in the following, when suitable with the purpose of clarity, we denote explicitly kets and bras in the physical Hilbert space as $|\, \rangle_p$ and $_p\langle \,|$ and analogously for the hidden Hilbert space $|\, \rangle_h$ and $_h\langle \,|$.

If $|+dx\rangle$ and $|-dx\rangle$ are normalized kets representing center of mass wawefunctions centered respectively at points $(+dx,0,0)$ and $(-dx,0,0)$, the state of the physical macroscopic ball
$$
|\psi \rangle_p =\frac{1}{\sqrt{2}} (|+dx\rangle+|-dx\rangle)
$$
corresponds to the metastate \cite{mypapers1}
$$
|\psi\rangle_m=\frac{1}{2}(|+dx\rangle>_p +|-dx\rangle_p)(|+dx\rangle_h+|-dx\rangle_h),
$$
whose (pure) meta density operator is
\begin{eqnarray}
\rho_m &=&\frac{1}{4}(|+dx\rangle_p+|-dx\rangle_p) (|+dx\rangle_h+|-dx\rangle_h)\cdot \nonumber \\
&& \;\; \left(_p\langle+dx|+_p\langle-dx|\right) \left( _h\langle+dx|+_h\langle-dx|\right). \nonumber
\end{eqnarray}
To obtain the physical density operator $\rho_p$ one has to trace out the hidden ball from the meta density operator $\rho_m$, i. e.
$$\rho_p = {\it Tr}[\rho_m]_h =_h\langle+dx|\rho_m|+dx\rangle_h + _h\langle-dx|\rho_m|-dx\rangle_h.
$$
 Let's consider now the time dependence due to the mutual gravitational interaction between the physical ball and its hidden partner observing that kets like $|+dx\rangle_p |+dx\rangle_h$ and $|-dx\rangle_p |-dx\rangle_h$ correspond to the physical ball and its hidden partner being perfectly superimposed except for the zero point motion that is to a very good approximation harmonic as can be checked by the mutual potential energy \cite{numericalsim} and corresponds to an irrelevant correction to the static gravitational energy for macroscopic bodies \cite{mypapers1,mypapers2}, whereas the kets $|+dx\rangle>_p |-dx\rangle_h$ and $|-dx\rangle_p |+dx\rangle_h$, assuming $dx \ge {\bold{r}}$, correspond to the two balls not overlapping at all.

As a result, taking account of the time evolution, we have the time dependent kets
$$
(|+dx\rangle_p |+dx\rangle_h) e^{-i \Omega_0 t}\, {\it and} \;\; (|+dx\rangle_p |-dx\rangle_h)e^{-i \Omega_1 t},
$$
where $\Omega_1=-\frac 12 (GM^2/(2dx))/h$ is the angular frequency corresponding to the mutual halved energy of the two non overlapping balls and $\Omega_0=-\frac 35 (GM^2/{\bold{r}})/h$ corresponds to the halved mutual energy of the two superimposed balls. By means of an elementary calculation one finds the time dependent physical density operator
\begin{eqnarray}
\rho_p(t)&=&\frac{1}{2}(|+dx\rangle\langle+dx|+|-dx\rangle\langle-dx|) + \nonumber \\
&&(|+dx\rangle\langle-dx|+|-dx\rangle\langle+dx|)\cos(\Omega t),\nonumber
\end{eqnarray}
where $\Omega=\Omega_0-\Omega_1$ is of the order of $(GM^2/{\bold{r}})/h$, which for a lead ball with ${\bold{r}}=0.1 m$,  $M=45 Kg$,  gives $\Omega$ of the order of $10^{26} sec^{-1}$. Assuming that the ray particles have a velocity $v=c/10$, along $1 m$ of their path they undergo the order of $10^{18}$ cycles of the oscillatory factor multiplying the coherencies, namely they feel only its vanishing average, by which for all practical practical purposes
\begin{equation}
\rho_p=\frac{1}{2}(|+dx\rangle\langle+dx|+|-dx\rangle\langle-dx|), \nonumber
\end{equation}
i. e. we get complete decoherence.
As a result the ray is always deflected with equal probability left or right independently of any measurement on the far away two state system. A detector on the $z$ axis beyond the ball never detects particles and so no SLC takes place.
\section{Conclusions}
The paper is neutral about the question whether SCG be meant to hold at fundamental level or as holding in peculiar circustances. The thesis expressed here is that in whatever sense SCG is considered, a possible instance of SLC is cured by NUNG. Although this was shown only for a particular instance of SLC, it is plausible that NUNG works in general since all of them depend on some sort of Schrodinger's cats that NUNG can decohere.

\section{Acknowledgements}
The author is grateful to Vittorio Del Duca, Joseph Quartieri and Mario Salerno for their unrelenting support.

\end{document}